\documentclass[pre,10pt,twocolumn,superscriptaddress,longbibliography,showpacs]{revtex4-1}
\usepackage{amsmath}
\usepackage{xcolor}
\usepackage{graphicx}

\newcommand{\W}{\mathrm{W}}
\newcommand{\T}{\mathrm{T}}
\newcommand{\G}{\mathrm{G}}

\renewcommand{\d}{\operatorname{d}}
\newcommand{\sign}{\operatorname{sign}}
\newcommand{\abs}[1]{\left| #1 \right|}
\begin{document}
\title{The transfer matrix approach to circular graphene quantum dots}

\author{H. Chau Nguyen}\email{chau@pks.mpg.de}
\affiliation{Max-Planck-Institut f\"{u}r Physik Komplexer Systeme, N{\"{o}}thnitzer Stra{\ss}e 38, 01187 Dresden, Germany}

\author{Nhung T. T. Nguyen}
\affiliation{Institute of Physics, VAST, 10 Dao Tan, Ba Dinh Distr., 118011 Hanoi, Vietnam}

\author{V. Lien Nguyen}
\affiliation{Institute of Physics, VAST, 10 Dao Tan, Ba Dinh Distr., 118011 Hanoi, Vietnam}
\affiliation{Institute for Bio-Medical Physics, 109A Pasteur, 1st Distr., 710115 Hochiminh City, Vietnam}

\begin{abstract}
We adapt the transfer matrix ($\T$-matrix) method originally designed for one-dimensional quantum mechanical problems to solve the circularly symmetric two-dimensional problem of graphene quantum dots. In similarity to one-dimensional problems, we show that the generalized $\T$-matrix contains rich information about the physical properties of these quantum dots. In particular, it is shown that the spectral equations for bound states as well as quasi-bound states of a circular graphene quantum dot and related quantities such as the local density of states and the scattering coefficients are all expressed exactly in terms of the $\T$-matrix for the radial confinement potential. As an example, we use the developed formalism to analyse physical aspects of a graphene quantum dot induced by a trapezoidal radial potential. Among the obtained results, it is in particular suggested that the thermal fluctuations and electrostatic disorders may appear as an obstacle to controlling the valley polarization of Dirac electrons.
\end{abstract}

\maketitle
\section{Introduction}
Transfer matrix ($\T$-matrix) is a classic quantum mechanics approach that is widely used to treat a variety of physical problems~\cite{Sanchez-Soto2012a}. Linearly relating  the parameters of the Schr\"{o}dinger waves in the two sides of a potential barrier, the $\T$-matrix contains a rich information of quantum characteristics of the potential examined. The effectiveness of the $\T$-matrix approach relies on its analytic simplicity and on the fact that $\T$-matrices can be easily multiplied when treating relatively complicated potential barriers. Exact expressions for the energy structure as well as the transport characteristics of semiconductor super-lattices that were derived by Esaki and Tsu~\cite{Tsu1973a} could be seen as a typical example of elegant successes of this approach.

As for the graphene, when charge carriers behave like the two-dimensional (2D) Dirac relativistic fermions, the $\T$-matrix approach has also been shown to be an effective approach. For the graphene nanostructures induced by one-dimensional (1D) potentials, such as the multi-barrier structures or the $n$-$p$-$n$-junctions, the $\T$-matrix calculations have been developed to study the energy spectrum~\cite{Nguyen2009a} as well as the dynamical characteristics ~\cite{Nguyen2009b}. In particular, the $\T$-matrix approach was successfully used to calculate the electronic band structure and the transport properties of various single-/bi-layer graphene superlattices induced by periodic electrostatic and/or magnetic potentials (see, for example,~\cite{Barbier2008a, Pham2014a} and references therein). Note that, traditionally, the $\T$-matrix approach was just suggested for (quasi) 1D potential problems.

The present work is devoted to another class of graphene nano-structures that are induced by a cylindrically symmetric potentials, known as circular graphene quantum dots (GQDs)~\cite{Apergel2010a}.  Experimentally, a circular GQD can be created using an appropriate circular top gate in the way as described in Ref.~\cite{Stander2009a}. Thanks to the fact that the gate potential can be tuned externally, such a gate-induced GQD can be easily controlled as regards its carrier density and effective radius. Theoretically, circular GQDs were often modelled by confinement potentials of either rectangular~\cite{Matulis2008a,Hewageegana2008a,Bardarson2009a, Recher2009a,Park2010a,Heinisch2013a,Guerrero-Becerra2014a,Schulz2014a} or power law forms~\cite{Chen2007a,Giavaras2009a,Giavaras2011a}. Then, by solving the Dirac-like equation for the chosen potential one obtained the dot energy spectrum and the associated quantities. It was shown that for the gapless pristine graphene in the absence of a magnetic field, due to the Klein tunneling, in general, it is not truly bound states but just quasi-bound ones with a finite trapping time that can be induced by an electrostatic confinement potential (see below, the text following Eq.~(\ref{eq: bound states}), for exceptional cases). An energy gap~\cite{Nguyen2009a,Recher2009a} and/or  a perpendicular magnetic field \cite{Chen2007a,Giavaras2009a,deMartino2007a} can enhance the trapping time of quasi-bound states (QBSs) and induce even the bound states. A smoothness of confinement potential was also shown to enhance the trapping time of QBSs. However, solving the Dirac-like equation with a smooth potential is often rather problematic.   

The purpose of this work is to extend the $\T$-matrix approach to study the electronic properties of circular GQDs induced by more general radial confinement potentials. For simplicity, our discussions are essentially limited to the case of zero magnetic field. Nevertheless, we briefly describe in an Appendix how to extend the approach to the case where a perpendicular homogeneous magnetic field is applied to the dot plane. 

Note that, in reality, a GQD with well-defined discrete energy levels can be created by cutting a structure with the desired geometry from a flake of graphene \cite{Todd2009a,Espinosa-Ortega2013a}. However, so far there is a serious problem in fabricating such GQDs with atomic precision termination, while it was shown that the electronic properties of these GQDs are quantitatively sensitive to their precise terminations~\cite{Espinosa-Ortega2013a}. From the future electronics application point of view it is desirable to find the way of creating GQDs by the confinement potentials so that the trapping time of localized states is long enough to satisfy the application requirements and the electronic properties of the structure can be  controlled externally.

The paper is organized as follows. Sec.~\ref{sec: general} presents the main results of the paper. It is there shown that for a very general class of circular GQDs, the bound and quasi-bound states spectral equations as well as the associated quantities, such as the local density of states and the resonance scattering characteristics, can all be expressed exactly in terms of the elements of the $\T$-matrix for the corresponding radial confinement potential. In Sec.~\ref{sec: examples} we show, as an example, the numerical solutions of the presented equations for the case of trapezoidal radial potential. Among the obtained results, it is in particular suggested that thermal fluctuations and/or electrostatic disorders may appear as an obstacle to controlling the valley polarization of Dirac electrons. While the paper is closed with a brief summary in Sec.~\ref{sec: conclusion}, the two Appendices are added to describe how the $\T$-matrix can be determined at some particular energies (A) and in the presence of a perpendicular magnetic field (B).     

\section{General consideration}
\label{sec: general}
Let us consider a single-layer circular GQD defined by the radial confinement potential $U(r)$ that is assumed to be smooth on the scale of the graphene lattice spacing. Using the units such that $\hbar=1$ and the Fermi velocity $v_F = 1$ (quasi-relativistic quantum units), the low-energy electron dynamics in this structure can be described by the 2D Dirac-like Hamiltonian
\begin{equation}
H= \vec{\sigma} \vec{p} + \nu \Delta \sigma_z + U(r),
\label{eq: H U}
\end{equation}
where $\vec{\sigma}=(\sigma_x,\sigma_y)$ are the Pauli matrices, $\vec{p}=-i (\partial_x, \partial_y)$ is the 2D momentum operator, $\nu$ is the valley index ($\nu = \pm$ for the valleys $K$ and $K'$, respectively) and $\Delta \sigma_z$ is the constant mass term~\cite{Neto2009a}. 

We look for the eigen-functions of the Hamiltonian~(\ref{eq: H U})  at energy $E$. Because of the cylindrical symmetry of $U(r)$, in the polar coordinates $(r, \phi )$ these eigen-functions can be found in the form
\begin{equation}
\Psi (r,\phi) = e^{i j \phi} 
\left( 
\begin{array}{cc}
e^{-i \phi/2} \chi_{A} (r) \\
e^{+i \phi/2} \chi_{B} (r)
\end{array}
\right), 
\label{eq: eigen j}
\end{equation}
where the total angular momentum  $j$ takes half-integer values and the radial spinor $\chi = (\chi_A,\chi_B)^{t}$ satisfies the following equation:
\begin{equation}
\left(
\begin{array}{cc}
U(r) - E + \nu \Delta & - i (\partial_r + \frac{j+\frac{1}{2}}{r}) \\
- i (\partial_r - \frac{j-\frac{1}{2}}{r}) & U(r)- E - \nu \Delta
\end{array}
\right) 
\left(
\begin{array}{c}
\chi_A(r)\\
\chi_B(r)
\end{array}
\right) 
= 0.
\label{eq: eigen chi U}
\end{equation}
This system of the two first order differential equations for the components $\chi_A$ and $\chi_B$ could be converted to a decoupled second order differential equation for either of these components. However, unless the potential $U(r)$ is simple enough, the resulting second order differential equations are often intractable. Nevertheless, we will show that the electronic characteristics of the circular GQDs described by the Hamiltonian~(\ref{eq: H U}) can be exactly expressed in terms of the elements of a $(2 \times 2)$ $\T$-matrix defined below.    

In order to define the expected $\T$-matrix, it should be noted that, in practice, we often have to deal with the confinement potentials $U(r)$ which are flat in the two limiting regions of small and large $r$, i.e.,
\begin{equation}
U(r)= 
\left\{ 
	\begin{array}{c}
	\mbox{$U_i$, \ \ \  $r \le r_i$},  \\
	\mbox{$U_f$, \ \ \ $r \ge r_f$}, \\
	\mbox{arbitrary, \ otherwise}.
	\end{array}
\right.
\label{eq: general potential}
\end{equation}
In these limiting regions, the eigenstates of  Hamiltonian (\ref{eq: H U}) can be found exactly. Indeed, we consider some region $r_a < r <  r_b$ where the potential $U(r)$ is constant, $U(r) = \bar{U}$. As is well-known~\cite{Recher2009a,Rubio2015a}, for $E \neq \bar{U} \pm \nu \Delta$ the 
general solution to Eq.~(\ref{eq: eigen chi U}) in this region can be written in terms of two independent integral constants $C = ( C^{(1)}, C^{(2)} )^t$:
\begin{equation}
\chi (r) = \W (\bar{U} , r) C ,
\label{eq: chi W}
\end{equation}
where the columns of the $\W$-matrix are the two independent basic solutions of Eq.~(\ref{eq: eigen chi U}),
\begin{equation}
\W (\bar{U} , r) = 
\left(
	\begin{array}{cc}
	J_{j-\frac{1}{2}} (q r) & Y_{j-\frac{1}{2}} (q r) \\
	i \tau J_{j+\frac{1}{2}} (q r) &  i \tau Y_{j+\frac{1}{2}} (q r)
	\end{array}
\right) .
\label{eq: W-matrix}
\end{equation}
Here  $J_{j\pm\frac{1}{2}}$ is the Bessel function of the first kind and  $Y_{j\pm\frac{1}{2}}$  is the Bessel function of the second kind~\cite{Abramowitz1972a}, $q = \sqrt{(E - \bar{U})^2 - \Delta^2}$ and $\tau = q / (E - \bar{U} + \nu \Delta )$. In the following, for definition, the integral constants $C = (C^{(1)}, C^{(2)})^t$ will be referred to as \emph{basic coefficients}. In the 1D problems, these basic coefficients can be interpreted as the coefficients of the forward and backward waves~\cite{Nguyen2009a,Nguyen2009b}. A similar interpretation can be introduced when the Hankel functions~\cite{Abramowitz1972a} are used to present the basic solutions $\W$~\cite{Abramowitz1972a}. In this paper, we however use the Bessel function representation for the sake of algebraic convenience.

A special care is needed in the case of energies $E \rightarrow \bar{U} \pm \nu \Delta$, when the basic solutions (\ref{eq: W-matrix}) become divergent. To avoid such a divergence, maintaining the matrix $\W$ as independent basic solutions, one has to properly adjust the regularization coefficients for the matrix elements in getting the correct limiting form of $\W$. To keep our discussions continuous, in the following we always assume that $E \ne \bar{U} \pm \nu \Delta$ and the case  $E \rightarrow \bar{U} \pm \nu \Delta$ will be discussed separately in Appendix~\ref{sec: zero-energy}.

We note that the basic coefficient $C$ can be considered as the spinor represented in a basis that depends on $r$ according to Eq.~(\ref{eq: W-matrix}). Then, Eq.~(\ref{eq: chi W}) actually describes a (non-unitary) basis transformation of the spinor. The advantage of using such a $r$-depending basis is that while $\chi(r)$ depends on $r$ explicitly, the wave coefficient $C$ is independent of $r$ in constant potential regions.

Now, the key feature of the differential Eq.~(\ref{eq: eigen chi U}) is that it is linear and homogeneous. Consequently, the two radial spinors at $r = r_1$ and $r = r_2$ should be linearly related by some matrix $\G (r_2,r_1)$,
\begin{equation}
\chi (r_2) = \G (r_2,r_1) \chi (r_1).
\label{eq: G-matrix}
\end{equation}
This relation holds for any $r_2 \ge r_1$, including the case of $r_1 \le r_i$ and $r_2 \ge r_f$ [see Eq.~(\ref{eq: general potential})]. Therefore, when we represent the spinors at $r_1 \le r_i$ and $r_2 \ge r_f$ by the basic coefficients $C_i$ and $C_f$ respectively, these basic coefficients should also be linearly related by some $\T$-matrix:
\begin{equation}
C_f = \T C_i .
\label{eq: T-matrix}
\end{equation} 
Note that the variable $r$ is entirely dropped out of this equation. Thus, in the context of the studied problem, the $\T$-matrix is defined as the matrix that maps the basic coefficients in the limiting region of small $r$ to those in the limiting region of large $r$. In fact, Eq.~(\ref{eq: T-matrix}) is a just basis transformation of Eq.~(\ref{eq: G-matrix}). From Eqs.~(\ref{eq: chi W}),~(\ref{eq: G-matrix}), and (\ref{eq: T-matrix}), we have the following elementary relation 
\begin{equation}
\T = \W^{-1} (U_f,r_2) \G (r_2,r_1) \W (U_i,r_1).
\label{eq: T-G}
\end{equation}
This equation, like Eq.~(\ref{eq: G-matrix}), holds for any $r_1 \le r_i$ and $r_2 \ge r_f$, including $r_1=r_i$ and $r_2=r_f$.

Equation~(\ref{eq: T-G}) provides a practical way to compute the $\T$-matrix for any radial potential $U(r)$ of Eq.~(\ref{eq: general potential}) via computing $\G(r_f,r_i)$. By inserting~(\ref{eq: G-matrix}) into~(\ref{eq: eigen chi U}), one finds an explicit differential equation for $\G(r_2,r_1)$, which resembles a dynamical equation in $r$-direction,
\begin{equation}
i\frac{\partial \G(r_2,r_1)}{\partial r_2}= \mathcal{H}(r_2) \G(r_2,r_1),
\end{equation}
with the formal Hamiltonian defined as
\begin{equation}
\mathcal{H} (r)= 
\left(
\begin{array}{cc}
i\frac{j-\frac{1}{2}}{r} & U(r) - E - \nu \Delta  \\
U(r)-E + \nu \Delta  & -i \frac{j+\frac{1}{2}}{r} 
\end{array} 
\right).
\end{equation}
This dynamical equation is to be solved for $\G(r_2,r_1)$ with the initial condition such that $\G(r_1,r_1)$ is the $(2 \times 2)$ identity matrix. Note that the formal Hamiltonian $\mathcal{H}(r)$ is not hermitian, and thus the dynamics is non-unitary. Moreover, $\mathcal{H}(r)$ at different $r$  generally do not commute with each other, rendering the dynamics analytically intractable. However, for the purpose of numerically calculating the $\T$-matrix, a simple numerical method for ordinary differential equations (ODEs) such as the Runge--Kutta method is sufficient~\cite{Press2002a}.  

Of particular importance is the case of one-step potential, $U(r)$ of Eq.~(\ref{eq: general potential}) with $r_i=r_f$. In this case, $\G(r_i,r_f)$ is simply the $(2 \times 2)$ identity matrix and we can easily write down the $\T$-matrix of Eq.~(\ref{eq: T-G}) explicitly,
\begin{widetext}
\begin{eqnarray}
\T && = \left[ \tau_f J_{j-\frac{1}{2}} (q_f r_f)  Y_{j+\frac{1}{2}} (q_f r_f) - \tau_f J_{j+\frac{1}{2}} (q_f r_f)Y_{j-\frac{1}{2}} (q_f r_f) \right]^{-1}  \nonumber \\
 && \times \left(
\begin{array}{cc}
 \tau_f  Y_{j+\frac{1}{2}} (q_f r_f)J_{j-\frac{1}{2}} (q_i r_i)- \tau_i Y_{j-\frac{1}{2}} (q_f r_f)J_{j+\frac{1}{2}} (q_i r_i)& \tau_f Y_{j+\frac{1}{2}} (q_f r_f) Y_{j-\frac{1}{2}} (q_i r_i) - \tau_i Y_{j-\frac{1}{2}} (q_f r_f)Y_{j+\frac{1}{2}} (q_i r_i) \\
- \tau_f J_{j+\frac{1}{2}} (q_f r_f)J_{j-\frac{1}{2}} (q_i r_i) + \tau_i  J_{j-\frac{1}{2}} (q_f r_f) J_{j+\frac{1}{2}} (q_i r_i) & - \tau_f J_{j+\frac{1}{2}} (q_f r_f) Y_{j-\frac{1}{2}} (q_i r_i)+ \tau_i J_{j-\frac{1}{2}} (q_f r_f) Y_{j+\frac{1}{2}} (q_i r_i) 
\end{array}
\right), \nonumber \\
\label{eq: explicit T}
\end{eqnarray}
\end{widetext}
where $q_{i(f)}$ and $\tau_{i(f)}$ are defined as in Eq.~(\ref{eq: W-matrix}): $q_{i(f)} = \sqrt{(E - U_{i(f)} )^2 - \Delta^2}$ and $\tau_{i(f)} = q_{i(f)} / (E - U_{i(f)} + \nu \Delta )$.

Being a seemingly simple mathematical consequence of the linearity and the homogeneity of the wave equations, the $\T$-matrix of Eq.~(\ref{eq: T-matrix}), as can be seen below, holds rich information on the characteristics of the energy spectrum of the system. In order to derive these characteristics, we are going to impose appropriate boundary conditions for the basic coefficients $C_i$ and $C_f$, which in turn lead to corresponding constraints on the elements of the $\T$-matrix itself. It should be noted immediately that in the limiting region of small $r$, the Bessel function of the first kind $J_{j\pm\frac{1}{2}}(q_i r)$ is regular, while  the Bessel function of the second kind $Y_{j\pm\frac{1}{2}}(q_i r)$ diverges. We should therefore set the condition $C_i \propto (1,0)^t $ for the basic coefficient in this region. We will first show that the localization behaviour of states is determined by the boundary condition for the basic coefficient $C_f$.
\subsection{Bound states} 
\label{sec: bound states}
For the bound states to emerge, the wave function should decay fast enough as $r$ increases. This happens only when the wave vector in the limiting region of large $r$, $q_f = \sqrt{( E - U_f )^2- \Delta^2}$, is imaginary, implying $-\Delta < E - U_f < \Delta$. Here, as mentioned above, we do not include the case of equalities, which may bring about a particular type of bound states (see also Appendix~\ref{sec: zero-energy} and references therein). Under this condition, although both Bessel functions  $J_{{j\pm\frac{1}{2}}} (q_f r)$ and $Y_{{j\pm\frac{1}{2}}} (q_f r)$ diverge as $r$ increases, the Hankel function of the first kind, $H^{+}_{{j\pm\frac{1}{2}}} (q_f r)= J_{{j\pm\frac{1}{2}}} (q_f r) + i Y_{{j\pm\frac{1}{2}}}(q_f r)$, decays exponentially. Thus, for the bound states to emerge, the appropriate boundary condition for the basic coefficient $C_f$ should have the form $C_f  \propto (1, i)^t $. With the boundary conditions for $C_i$ and $C_f$ just defined, Eq.~(\ref{eq: T-matrix}) leads to the following relation for the elements of the $\T$-matrix:
\begin{equation}
\T_{11} + i  \T_{21} = 0.
\label{eq: bound states}
\end{equation}
This is the general equation to determine the energy spectrum of all the bound states in the considered energy regions for a GQD induced by the potential of Eq.~(\ref{eq: general potential}). To obtain this energy spectrum we first have to calculate the $\T$-matrix in the way described above and then to solve Eq.~(\ref  
{eq: bound states}). In the particular case of one-step potentials, using the explicit $\T$-matrix of Eq.~(\ref{eq: explicit T}), we can easily recover the bound state spectral equation reported in Refs.~\cite{Recher2009a,Rubio2015a} for the GQD induced by a rectangular potential. 
\subsection{Quasi-bound states}

For $| E - U_f  | >  \Delta$, the wave vector in the limiting region of large $r$, $q_f$, is always real and the corresponding states cannot be truly bound. However, carriers may be temporally trapped at these states with some finite life-time. As mentioned above, such states are often referred to as the QBSs.  Each QBS can be characterized by a complex energy $E = \Re(E) + i \Im(E)$ with $\Im(E)<0$. The real part of this energy, $\Re(E)$, defines the position of the QBS (i.e., the resonant level), while the imaginary part, $\Im(E)$, causes the probability density of the QBS to decay over time $t$ as $ \propto e^{2 \Im(E) t}$. This implies that $| \Im(E) |$ is a measure of the resonant level width and its inverse is a measure of the carrier life-time at the QBS,  $\tau_0 \propto 1/ (2 | \Im(E) |)$. 

Actually, the way we determined the spectral equation for bound states, Eq.~(\ref{eq: bound states}), can be easily extended to find the spectrum of QBSs. Indeed, as well-known~\cite{Chen2007a,Nguyen2009a}, the reasonable boundary condition for QBSs is that far from the origin the wave function should be an out-going wave. Letting  $s = \sign(E - U_f )$, it is easy to see that the wave function $(H^{s}_{{j-\frac{1}{2}}} (q_f r), i \tau_f H^{s}_{{j+\frac{1}{2}}} (q_f r))^t $ with $H^{s}_{{j\pm\frac{1}{2}}} (q_f r) = J_{{j\pm\frac{1}{2}}} (q_f r) + is Y_{{j\pm\frac{1}{2}}} (q_f r)$ describes such an out-going wave. This can be confirmed by examining the current density of the radial wave function in the limiting region of large $r$ using the well-known asymptotic forms of the Hankel functions~\cite{Abramowitz1972a}. With the wave-function identified, in terms of the basic coefficients, it appears that the appropriate boundary condition for QBSs takes the simple from: $C_f \propto (1, is)^t$. Using this $C_f$ and the boundary condition for $C_i$ defined above, Eq.~(\ref{eq: T-matrix}) results in the general equation for determining the QBSs spectrum in circular GQDs:
\begin{equation}
\T_{11} + i s \T_{21} = 0.
\label{eq: quasi-bound states}
\end{equation}
Note that, to our best knowledge, the QBSs in circular GQDs were often identified by either  numerically fitting asymptotic boundary conditions~\cite{Chen2007a}, or intuitively analysing the behaviour of the local density of states~\cite{Matulis2008a,Masir2009a}. Equation~(\ref{eq: quasi-bound states}) provides an alternative way to solve the problem, making it more definite and rather simple algebraically. In fact, this equation is in the same spirit as the equation suggested sometime ago for the QBSs in a 1D potential~\cite{Nguyen2009b}.

\subsection{Density of states}
The local density of states (LDOS) for unbound states, as defined in~\cite{Matulis2008a}, can also be easily expressed in terms of the $\T$-matrix of the radial confinement potential. Note that for unbound states the wave functions are not normalizable and the usual definition of LDOS~\cite{Davies1998a} should be used with care. Following~\cite{Matulis2008a}, we image that the considered GQD is entirely embedded in a large graphene disc of radius $R$, with the center of this disc  coincides with that of the GQD. States are then bound within the large graphene disc, and the level spacing can be estimated to be $\Delta E= \pi/R$~\cite{Matulis2008a}. The LDOS of the GQD is proportional to both the level density and the probability for the electron at that energy level to be inside the dot. For a wave function with basic coefficients $C_i = (F, 0)^t$ and $C_f = (P, Q)^t$, the latter is proportional to $\abs{F}^2 / \abs{N}^2$, where $N$ is the normalization factor of the wave function, which in turn can be estimated to be $\abs{N}^2 \propto (\abs{P}^2 + \abs{Q}^2) R/\abs{E}$~\cite{Matulis2008a}. Overall, this gives the formula for the LDOS: $\rho^{(j)} (E) \propto   \abs{E} \abs{F}^2  / ( \abs{P}^2+\abs{Q}^2 )$.
 
In order to get the LDOS in terms of the $\T$-matrix, we can use the relation~(\ref{eq: T-matrix}) to show that $\abs{F}^2/(\abs{P}^2+\abs{Q}^2) = 1 / ( \abs{T_{11}}^2+\abs{T_{21}}^2)$. Thus, for a given angular momentum $j$ and a given valley index $\nu$, the LDOS around the circular GQD can be calculated in terms of the $\T$-matrix as
\begin{equation}
\rho^{(j)} (E) \propto \frac{\abs{E}}{\abs{T_{11}^{(j)}}^2 + \abs{T_{21}^{(j)}}^2},
\label{eq: dos}
\end{equation}
where the superscript $(j)$ is added to explicitly indicate the $j$-dependence of the quantity calculated.  Summing (\ref{eq: dos}) over all angular momenta, we obtain the total LDOS,
\begin{equation}
\rho(E)= \sum_{j=-\infty}^{+\infty} \rho^{(j)} (E).
\label{eq:tldos}
\end{equation}
It is easy to show that these general expressions, Eqs.~(\ref{eq: dos}) and (\ref{eq:tldos}), directly reduce to the corresponding ones given in Ref.~\cite{Matulis2008a} for circular GQDs with a rectangular confinement  potential.
 
\subsection{Scattering coefficients}
\label{sec: scattering}
The scattering states are those with the asymptotic wave functions far from the origin being a superposition of an in-coming plane wave and an out-going (scattering) circular wave~\cite{Sakurai1994a}. Thus, for $r > r_f $, we write
\begin{equation}
\Psi_f (r,\phi) = \Psi_f^{(i)} (r, \phi) + \Psi_f^{(o)} (r, \phi),
\end{equation}
where the first and the second terms in the right-hand-side are the in-coming plane wave and out-going circular wave, respectively. The in-coming wave function $\Psi_f^{(i)}(r, \phi )$ is assumed to propagate along the $x$-direction with positive current density,
\begin{equation}
\Psi_f^{(i)} (r,\phi)= e^{isq_f r \cos \phi} 
\left( 
\begin{array}{c}
1 \\
\frac{s q_f}{E-U_f+ \nu \Delta}
\end{array}
\right),
\label{eq:plane-wave}
\end{equation}
where $q_f$ and $s$ have already been defined above. Note that for the electron to be propagated at large $r$, the energy should not be in the gap, $\abs{E-U}>\Delta$. Using the Jacobi--Anger identity~\cite{Cuyt2008a}, the plane wave function of Eq.~(\ref{eq:plane-wave}) can be decomposed into the eigen-functions of the angular momentum  as
\begin{equation}
\Psi_f^{(i)}(r,\phi)=\sum_{j=-\infty}^{+\infty} (is)^{j-\frac{1}{2}}  e^{ij{\phi}}
	\left( 
	\begin{array}{c}
	e^{-\frac{i}{2}\phi} \ J_{j-\frac{1}{2}}(q_f r) \\
	e^{+\frac{i}{2}\phi} \ i\tau_f J_{j+\frac{1}{2}}(q_f r)
	\end{array}
	\right),
\end{equation}
with $\tau_f$ also already defined. 

The scattering wave can be also expanded in the out-going waves of different angular momenta,
\begin{equation}
\Psi_f^{(o)} (r,\phi)=\sum_{j= -\infty}^{+\infty} a^{(j)} (is)^{j-\frac{1}{2}}  e^{ij{\phi}}
	\left( 
	\begin{array}{c}
	e^{-\frac{i}{2}\phi} \ H^{s}_{{j-\frac{1}{2}}}(q_f r)  \\
	e^{+\frac{i}{2}\phi} \ i\tau_f H^{s}_{{j+\frac{1}{2}}}(q_f r)
	\end{array}
	\right),
	\label{eq: out-going}
\end{equation}
where $a^{(j)}$ are regarded as scattering coefficients~\cite{Heinisch2013a,Schulz2014a,Schulz2015a}.

For $r<r_i$, similarly, the wave function can be decomposed into a linear combination of the wave functions of different angular momenta. Noting that to ensure the regularity of the wave function at the origin, the Bessel functions of the second kind are necessarily absent from this decomposition, one has
\begin{equation}
\Psi_i (r,\theta) =\sum_{j=-\infty}^{+\infty} c^{(j)} (is)^{j-\frac{1}{2}}  e^{ij{\phi}}
	\left( 
	\begin{array}{c}
	e^{-\frac{i}{2}\phi} \ J_{{j-\frac{1}{2}}}(q_i r) \\
	e^{+\frac{i}{2}\phi} \ i\tau_{i} J_{{j+\frac{1}{2}}}(q_i r)
	\end{array}
	\right),
\end{equation}
with $q_i$ and $\tau_{i}$ defined before and $c^{(j)}$ being some coefficients.

Further, since the basic coefficients in the two limiting regions, $r \le r_i$ and $r \ge r_f$, should be related to each other by the $\T$-matrix as in Eq.~(\ref{eq: T-matrix}), we find
\begin{equation}
	\left( 
	\begin{array}{c}
	a^{(j)}+1 \\
	isa^{(j)}
	\end{array}
	\right)
=\T^{(j)}
\left( 
	\begin{array}{c}
	c^{(j)} \\
	0
	\end{array}
	\right),
	\label{eq: amplitude equation}
\end{equation}
where the superscript $(j)$ is again introduced to indicate the $j$-dependence of $\T$-matrix. Solving Eq.~(\ref{eq: amplitude equation}) gives the scattering coefficients in terms of the $\T$-matrix elements:
\begin{equation}
	a^{(j)}=\frac{-isT^{(j)}_{21}}{T^{(j)}_{11}+isT^{(j)}_{21}}.
	\label{eq: scattering coefficients}
\end{equation}
Now it is important to note that for an unbound eigen-function of real energy, to ensure the probability current conservation, it requires that the coefficients for the total out-going waves and the total in-going waves should be equal in modulus~\cite{Sakurai1994a}, 
\begin{equation}
\abs{T_{11}^{(j)} - is T_{21}^{(j)}}  = \abs{T_{11}^{(j)} + is T_{21}^{(j)}}.
\end{equation}
This implies that the scattering coefficients $a^{(j)}$ can be represented in terms of the so-called scattering phase-shifts~\cite{Sakurai1994a,Masir2011a},
\begin{equation}
a^{(j)}= \frac{1}{2} \left(e^{-i 2 \delta^{(j)}} -1\right),
\end{equation}
where 
\begin{equation}
\delta^{(j)} = \frac{1}{2} \arg \left( \frac{T_{11}^{(j)} + is T_{21}^{(j)}}{T_{11}^{(j)} - is T_{21}^{(j)}}\right).
\label{eq: phase-shift}
\end{equation}

The differential scattering cross section, defined as the ratio of the probability flux of the out-going wave per unit angle to the probability flux of the in-coming wave per unit length~\cite{Sakurai1994a,Masir2011a}, can be found as
\begin{equation}
\frac{\d \sigma }{\d \phi} = \frac{2}{\pi q_f} \abs{\sum_{j=-\infty}^{+\infty} a^{(j)} e^{j \phi}}^2.
\label{eq: scattering cross section}
\end{equation}
By integrating this expression over $\phi$, one finds the total scattering cross section,
\begin{equation}
\sigma = \frac{4}{q_f} \sum_{j=-\infty}^{+\infty} \sin^2 \delta_j.
\end{equation}  

\bigskip
Thus, for circular GQDs with an arbitrary radial confinement potential of  Eq.~(\ref{eq: general potential}), we have shown that the bound states as well as the QBSs spectra and the associated quantities such as the LDOS and the scattering coefficients can all be exactly expressed in terms of $\T$-matrix elements. Equations~(\ref{eq: bound states}), (\ref{eq: quasi-bound states}), (\ref{eq: dos}), and (\ref{eq: scattering coefficients}) are the key results of the present work. In particular cases, when the eigenstates of Hamiltonian (\ref{eq: H U}) can be found analytically (e.g., for a rectangular potential $U(r)$), these equations are exactly reduced to the corresponding expressions reported in various references. Generally, the $\T$-matrix can be calculated numerically. In the next section, as an example, we present numerical results obtained in the case of trapezoidal radial confinement potential.

\section{Example: Trapezoidal radial potential induced GQDs}
\label{sec: examples}
As a demonstration for the studies presented in the previous section, we  consider a circular GQD induced by the radial potential of Eq.~(\ref{eq: general potential}) with: $U_i = U_0$, $r_i = (1-\alpha) L$, $U_f = 0$, $r_f = (1+\alpha) L$ and $U(r) = U_i + \frac{r - r_i }{ r_f - r_i }(U_f - U_i )$ for $r_i < r < r_f$. So, the considered confinement potential has a trapezoidal shape that is characterized by three parameters: the potential height $U_0$, the dot effective radius $L$, and the smoothness $\alpha$ that ranges from $0$ to $1$. In the limiting case of  $\alpha = 0$, this potential is just the most studied rectangular one. The 1D trapezoidal potential are often used to describe the gate-induced graphene $n$-$p$-$n$-junctions \cite{Huard2007a,Sonin2009a}.   

\begin{figure}[!hbt]
\centering
\includegraphics[width=0.45\textwidth]{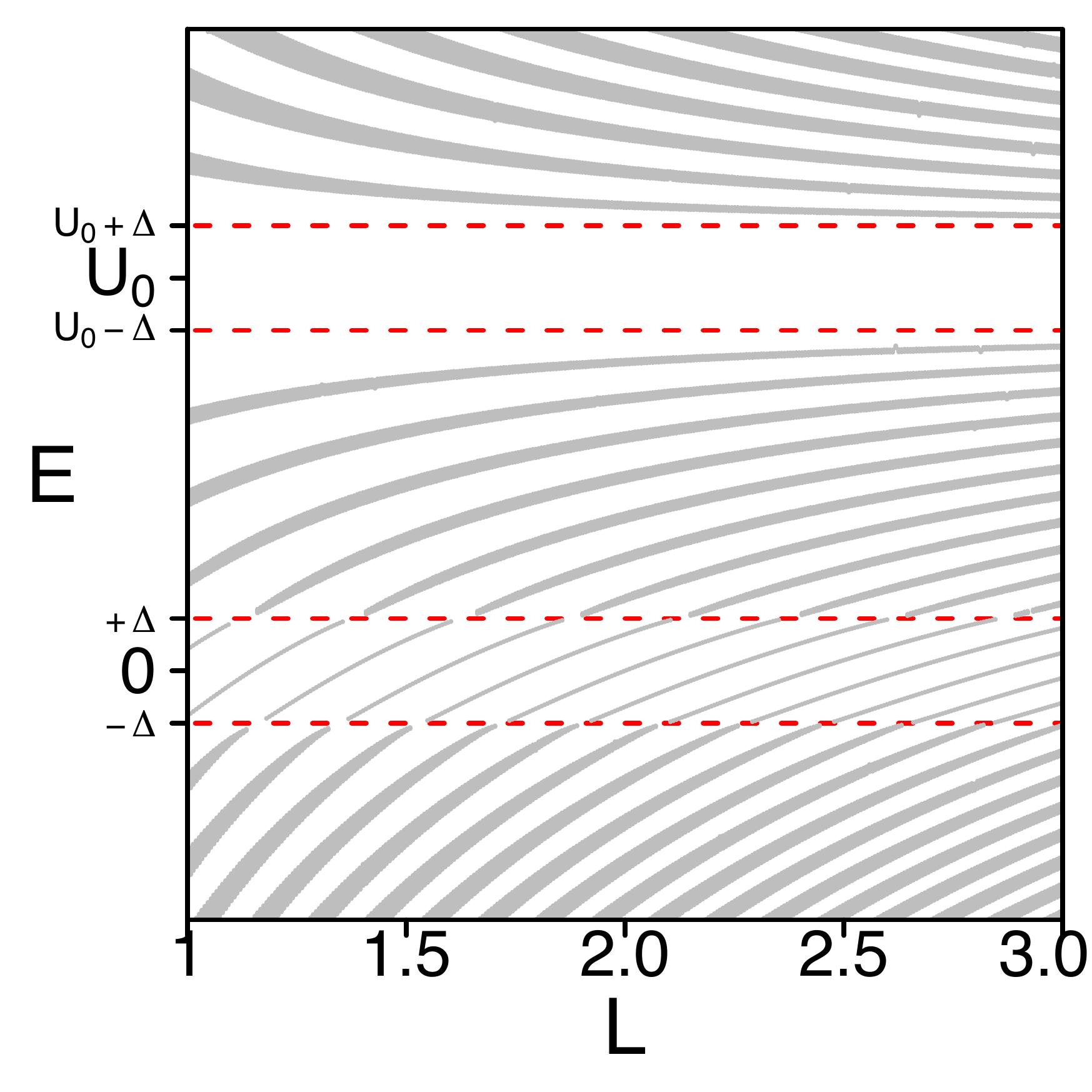}
\caption{(Colour online) Spectrum of bound states calculated from  Eq.~(\ref{eq: bound states}) and QBSs 
       from Eq.~(\ref{eq: quasi-bound states}) for a GQD induced by the trapezoidal radial potential
       of $U_0 = 15$ and $\alpha=0$. The lines represent the level
       positions, plotted versus the dot effective radius L, while the
       thickness of these lines represents the corresponding level widths.
       Data are shown for $\nu=+$, $j = \frac{3}{2}$ and $\Delta=2$.}
\label{fig: spectrum L}
\end{figure}

\begin{figure}[!hbt]
\centering
\begin{minipage}{0.47\textwidth}
\begin{center}
\includegraphics[width=\textwidth]{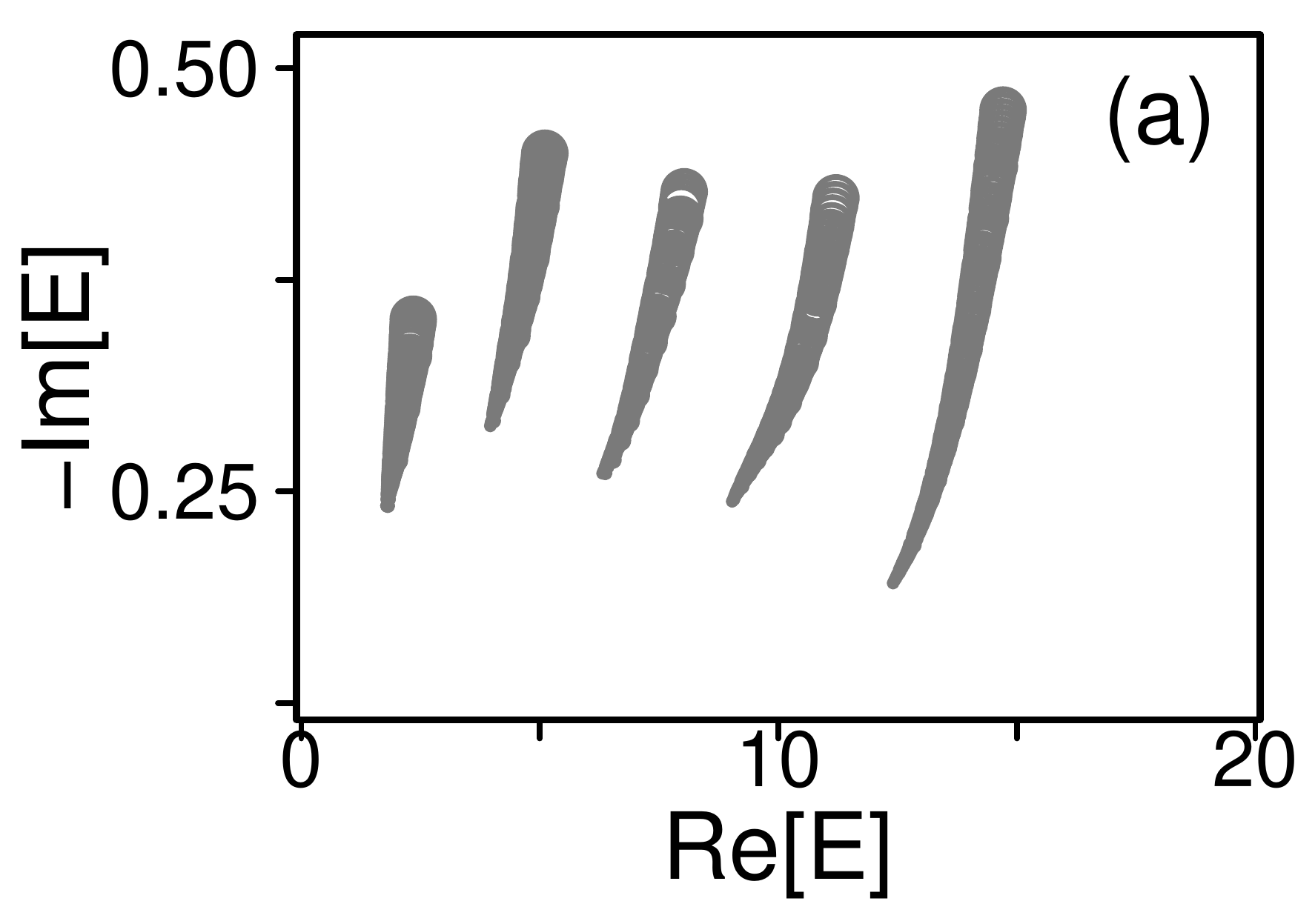}
\end{center}
\end{minipage}\\
\begin{minipage}{0.47\textwidth}
\begin{center}
\includegraphics[width=\textwidth]{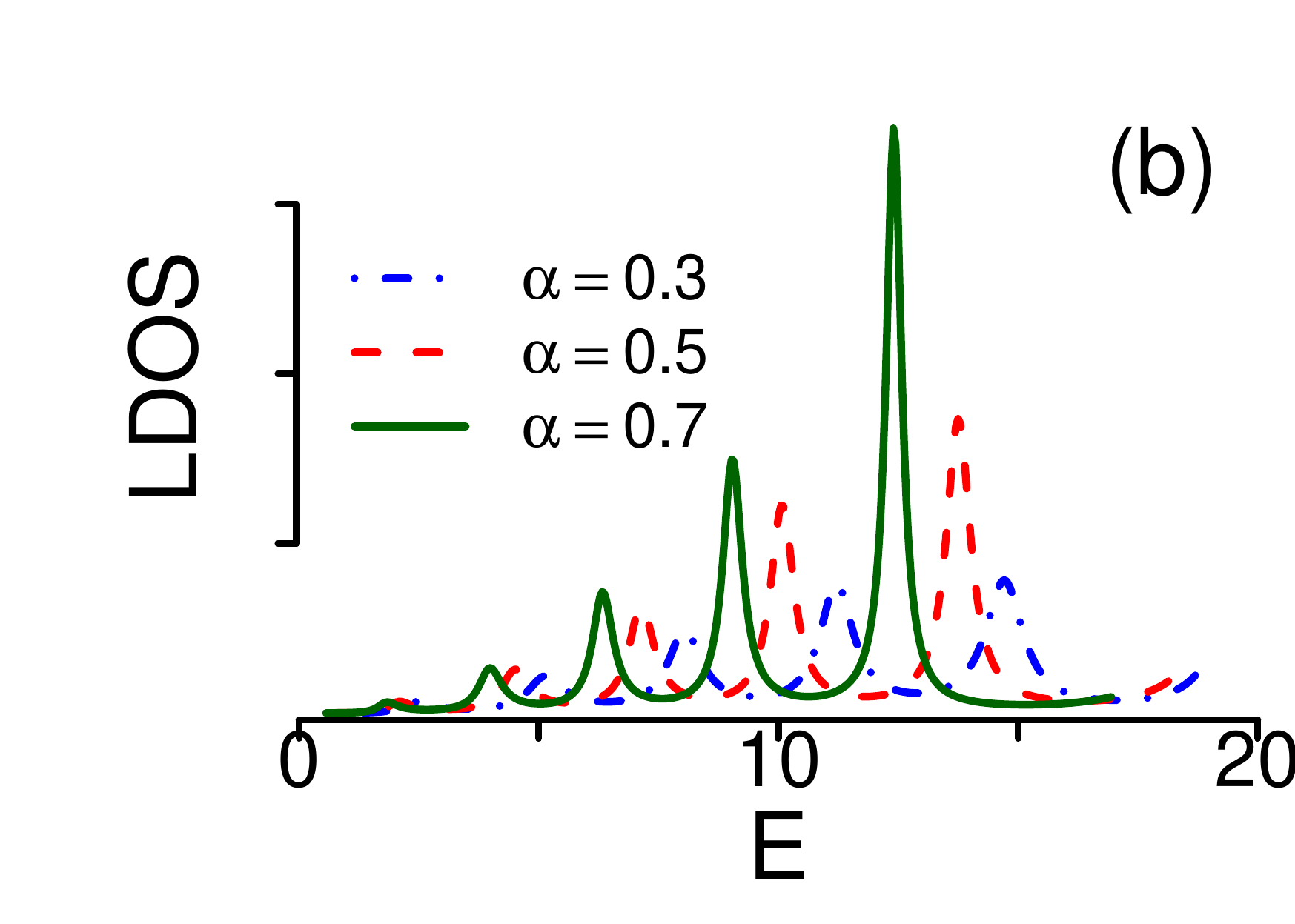}
\end{center}
\end{minipage} 
\caption{(Colour online) QBS spectra $(a)$ and LDOS $(b)$ of a GQD induced by the trapezoidal radial potential  of  $L=1$ and $U_0=20$ are presented for $\nu=+$, $j = \frac{3}{2}$ and various $\alpha$. In $(a)$: 5 curves correspond to 5 QBS levels, each describing how the QBS energy ($\Im(E)$ and $\Re(E)$) changes as $\alpha$ varying regularly from 0.3 (top) to 0.7 (bottom), correspondingly, from larger point-sizes to smaller point-sizes. In $(b)$: LDOS (in arbitrary unit) is shown for the three spectra with $\alpha$ given in the figure.}
\label{fig: spectrum a}
\end{figure}

For given values of potential parameters as well as the angular momentum $j$, we first calculate the $\T$-matrix for the potential under study. In the case of $\alpha \ne 0$, the calculation of the $\T$-matrix requires solving the ODE~(\ref{eq: T-G}) numerically for the matrix $\G(r_i,r_f)$ with the Runge--Kutta method. Substituting the obtained $\T$-matrix elements into Eqs.~(\ref{eq: bound states}), (\ref{eq: quasi-bound states}), (\ref{eq: dos}), and (\ref{eq: scattering coefficients}), and solving these equations, we respectively obtain the energy spectra, the associated LDOS, and the scattering coefficients~\footnote{For the indicated parameters, the Runge--Kutta method with about $1024$ steps gave the typical accuracy of $10^{-5}$ for the elements of the $T$-matrix. The numerical solutions of Eqs.~(\ref{eq: bound states}) and ~(\ref{eq: quasi-bound states}) presented in Fig.~\ref{fig: spectrum L} and Fig.~\ref{fig: spectrum a} (a) were obtained at the effective resolution of at least $4026$ grid-points in each dimension. Bessel functions were computed using the corresponding subroutines from Ref.~\cite{Zhang1996b}}. Such calculations can be carried out for various values of the potential parameters and the angular momenta. As an example, some of the obtained results are presented in Figs.1-4.

Note that we still use the quasi-relativistic quantum units ($\hbar=1$, $v_F=1$), so the dimension of energy is inverse of the length. For a comparison, to describe the usual experimental values of $L$ and $U_0$ ($L$ is of the order of $100$ nm and $U_0$ is of the order of $130$ meV), we choose $L$ to be about $1$ and $U_0$ to be  about $20$.

We first set $\alpha=0$ and study the spectra of bound states and QBSs as $L$ changing from $1$ to $3$. Obtained results are shown in Fig.~\ref{fig: spectrum L}. The limiting lines $E = \pm \Delta$ and $E = U_0 \pm \Delta$ define qualitatively different energy regions. The region $U_0 - \Delta  \le E \le U_0 + \Delta $ appears as a gap, where there exists neither bound states nor QBSs. On the other hand, the states in the region of energies $- \Delta  \le E \le +\Delta $ are truly bound, while those outside these regions are QBSs. For the QBSs presented, the thickness of the lines represents the corresponding level widths.  When $L$ increases, starting from the low energy region ($E <  -\Delta$), the QBS-levels gradually rise to approach the boundary at $E=-\Delta$, and, at the same time, their widths gradually narrow to vanish at this boundary. Throughout the region $-\Delta<E<+\Delta$, the states are truly bound with zero level widths. At the opposite boundary $E=+\Delta$ the states are again converted to QBSs. So, there may observe a continuous QBS - bound state - QBS transition in the energy spectra of circular GQDs  as the dot radius $L$ varies. Note that,  in the case of zero-gap, $\Delta = 0$, the bound states region actually collapses into the line $E = 0$ (That is why these states have been referred to as zero-energy ones~\cite{Hewageegana2008a,Downing2011a}). At very large $L$, all levels converge to the two boundaries $E= U_0 \pm \Delta$ that describe the limiting case when a homogeneous potential of $U_0$ is applied on the entire graphene sheet.

In the gapless case, $\Delta = 0$, all the states other than zero-energy ones  are just QBSs. In this case, the QBSs with energies in the region $0<E<U_0$ tend to have the level widths narrower than that for the QBSs with energies outside this region. It was suggested that the level widths of these QBSs can also be tuned by varying the smoothness of the confinement potential $\alpha$~\cite{Chen2007a}. Fig.~\ref{fig: spectrum a} $(a)$ shows how the complex energies of five different QBSs change as the smoothness $\alpha$ varies from $0.3$ to $0.7$ (correspondingly, point-sizes gradually decrease). Obviously, for any QBS under study, with increasing potential smoothness $\alpha$, while the real part of the energy $\Re(E)$ just changes slightly, the imaginary part $\Im(E)$ decreases substantially. This result is in a good agreement with those reported for 1D potentials~\cite{Nguyen2009a} and 2D power law potentials~\cite{Chen2007a}.

Next, we show in Fig.~\ref{fig: spectrum a}~$(b)$ the LDOSs (in arbitrary unit) for the three spectra with the $\alpha$-values examined in Fig.~\ref{fig: spectrum a}~$(a)$. Evidently, there is  a good agreement between the positions of QBSs in $(a)$ and the corresponding resonant peaks of LDOS in $(b)$. Moreover, the imaginary parts of the QBS energies represent the widths of the corresponding LDOS peaks quite well. Thus, our results qualitatively demonstrate the correspondence between the QBSs and the LDOS peaks. In fact, the LDOS has already been used to determine QBSs indirectly~\cite{Masir2009a}. Quantitatively, it should however be noted that for very broad LDOS peaks, such as those at $E \approx 1$ in Fig.~\ref{fig: spectrum a}~$(b)$, the peak width may not correctly describe the life-time of the corresponding QBS. 

\begin{figure}[!hbt]
	\centering
	\includegraphics[width=0.45\textwidth]{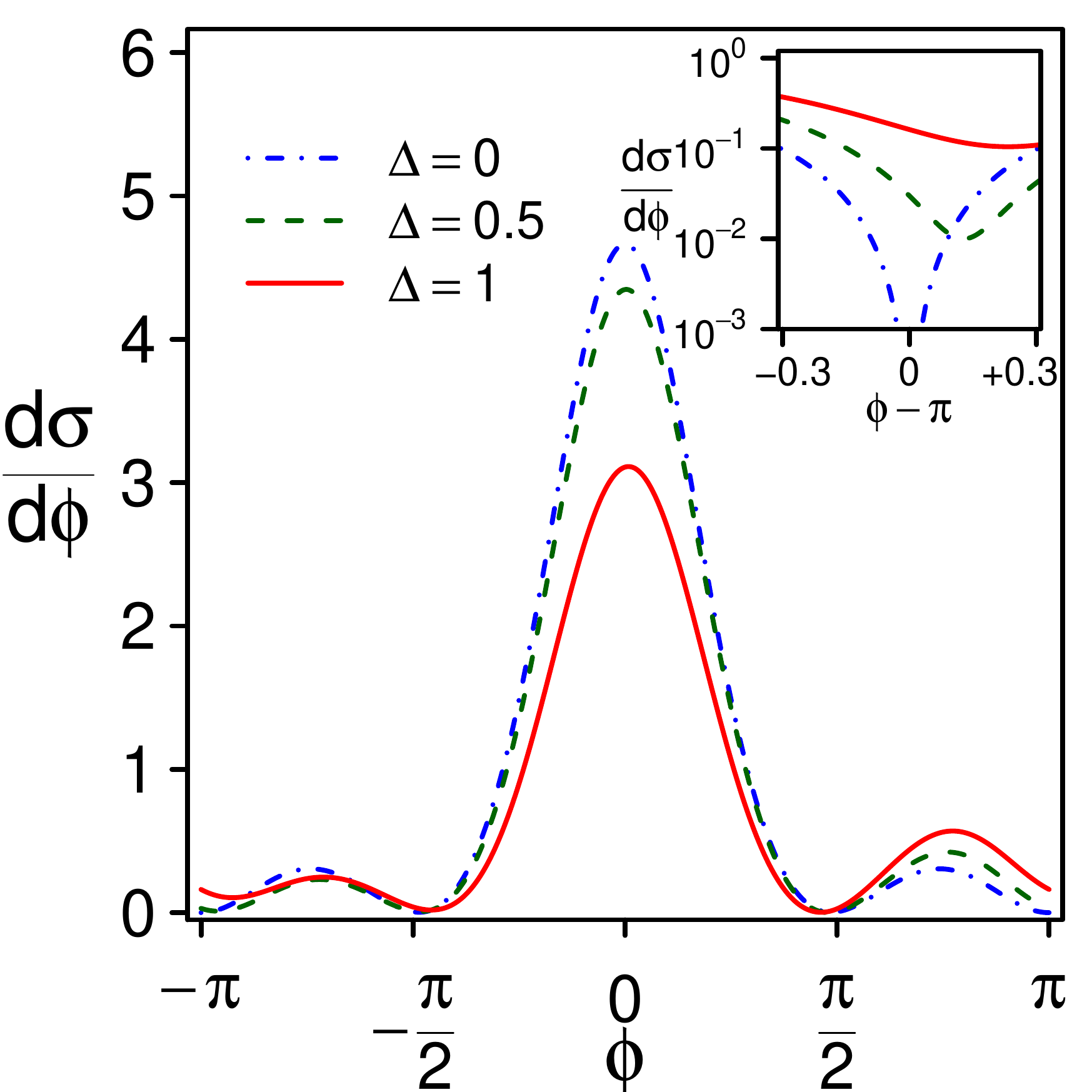}
	\caption{(Colour online) Low-energy differential scattering cross section is plotted as a function of scattering angle $\phi$ for the trapezoidal radial potential of $U_0=15$, $L=1$,  and $\alpha=0$ in three cases of $\Delta$: $0$ (dash-dotted line), $0.5$ (dashed line), and $1$ (solid line). The inset zooms in the region of scattering angle around $\pi$. Data are shown for $E=2$ and $\nu=+1$.}
	\label{fig: scattering}
\end{figure}

To illustrate the $\T$-matrix-based scattering formalism developed in subsection~\ref{sec: scattering}, we calculate the low-energy differential scattering cross section $\d \sigma / \d \phi $ for the trapezoidal potential of $U_0=20$ and $L= 1$ ($\alpha$ is set to be zero for simplicity). In Fig.~\ref{fig: scattering}, obtained results of $\d\sigma / \d\phi $ are presented as a function of the scattering angle $\phi$ in three cases: $\Delta = 0$ (gapless), $0.5$, and $1$ (finite gap). In the gapless case (dash-dotted line), the differential scattering cross section vanishes at $\phi = \pm \pi$ (Fig.~\ref{fig: scattering}, inset), showing the undoubted effect of the Klein tunnelling. In the two cases of finite gap, on the contrary,  $\d\sigma / \d\phi $ is always finite, implying an unavoidable presence of the back-scattering.

Besides, the two curves of finite gap (solid and dashed) in Fig.~\ref{fig: scattering} clearly show an asymmetrical behaviour with respect to the sign of $\phi$. A similar asymmetry has been discussed in the context of scattering of Dirac electrons by the so-called mass-barriers in Ref.~\cite{Masir2011a}. Note that by the reflection symmetry, $j \rightarrow -j$, $\nu \rightarrow -\nu$, electrons with opposite valley indices will scatter as if reflected along $\phi=0$, so no Hall-like voltage can be expected unless  the injected current is valley-polarized. Nevertheless, with an unpolarised current, electrons of different valley indices are expected to accumulate on opposite edges of the graphene sample in the way similar to the spin Hall effect~\cite{Sinova2015a}. The valley-dependent asymmetric scattering was suggested to be used for the valley filtering purpose~\cite{Masir2011a}.  

\begin{figure}[!hbt]
	\centering
	\includegraphics[width=0.45\textwidth]{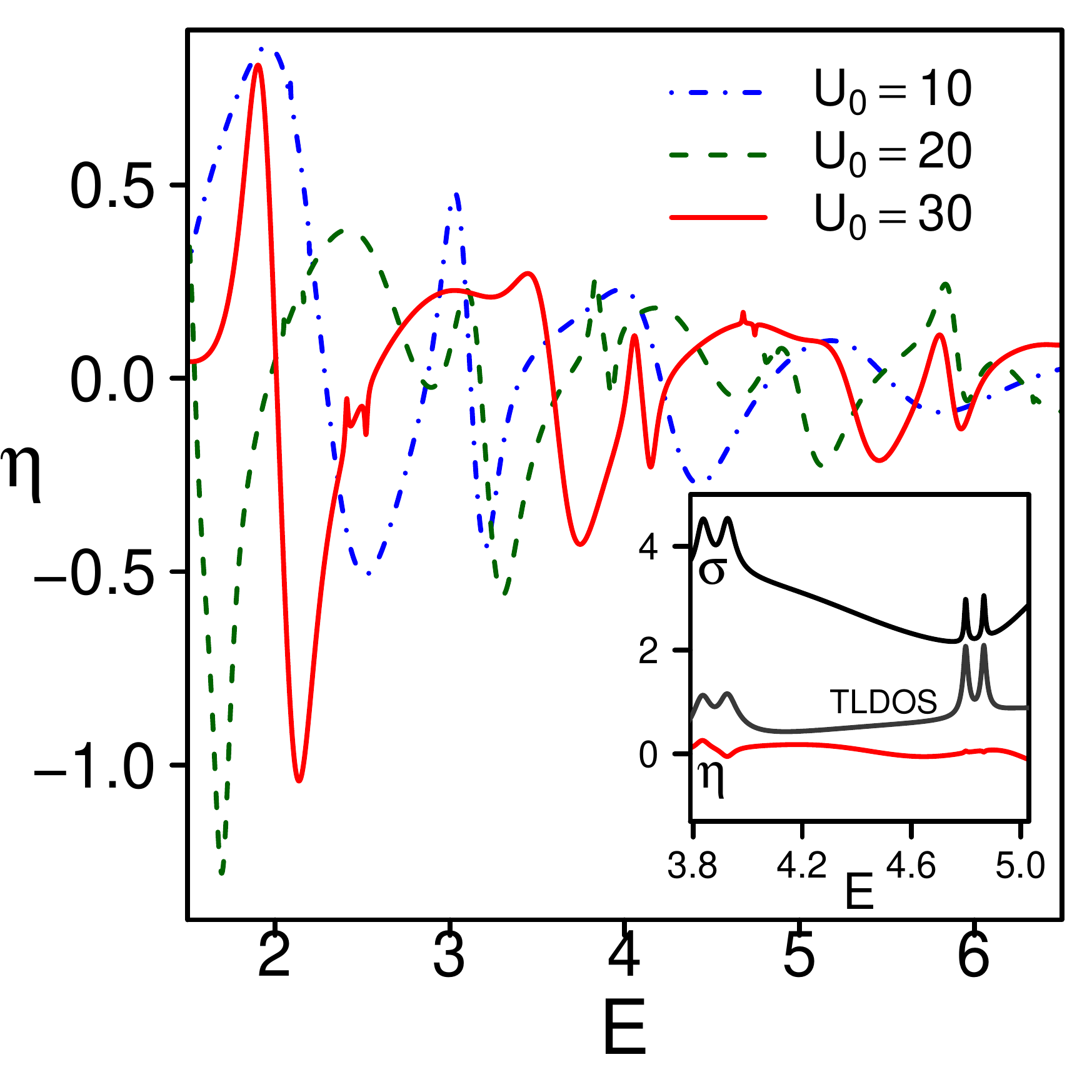}
	\caption{ (Colour online) Transverse scattering cross section $\eta$ as a function of the incident energy $E$  for potentials of $L=1$, $\alpha=0$, and various $U_0$:   $10$ (dash-dotted line), $20$ (dashed line), and $30$ (solid line). The inset zooms in a small region of energy (for $U_0=30$), where the total scattering cross section (labeled $\sigma$) and the (total) local density of states (labelled TLDOS) are also plotted for a comparison. [Note that the TLDOS (defined up to a constant factor) was rescaled to fit the figure.] Data are shown for $\Delta= 0.5$ and $\nu=+1$.}
	\label{fig: eta}
\end{figure}

Further, to learn if the examined electrostatic potential can support to control the valley polarisation of Dirac electrons like the mass potential does ~\cite{Masir2011a}, we calculate the transverse scattering cross section defined as
\begin{equation}
\eta= \int_{-\pi}^{+\pi}  \d \sigma (\phi) \sin \phi.
\end{equation} 
Calculations have been performed for potentials of $L = 1$, $\alpha = 0$, and different $U_0$. Obtained results for $\eta$ are plotted as a function of the incident energy $E$ in Fig.~\ref{fig: eta}, where the three curves are different in $U_0$: $U_0 = 10$ (dash-dotted line), $20$ (dashed line), and $30$ (solid line). Remarkably, $\eta$ strongly fluctuates, changing its sign in a complicated way, depending on both $E$ and $U_0$. Consequently, the transverse scattering cross sections of valley-polarized electrons of slightly different energies (e.g., due to thermal fluctuations), or from slightly different potentials, might compensate each other, resulting in a vanishing net transverse scattering cross section. This is very different from the scattering of Dirac electrons by a mass-barrier studied in Ref.~\cite{Masir2011a}, where it was shown that the transverse scattering cross section generally keeps its sign unchanged as the energy of electron varies. Given the fact that an energy gap in graphene is often induced by an underlying substrate~\cite{Neto2009a,Guinea2010a}, a mass-barrier is likely to be accompanied by electrostatic disorders. Thus, although a more quantitative study is needed, we speculate that the electrostatic disorders and/or the thermal fluctuation may appear as an obstacle to controlling the valley polarization of Dirac electrons and, therefore, to observing the associated zero-field Hall and the valley filtering effects~\cite{Masir2011a,Guinea2010a} .

Finally, to gain some insight into the discussed fluctuating behaviour of the transverse scattering cross section $\eta$ observed in Fig.~\ref{fig: eta}, in the inset to this figure we compare three quantities, $\eta$, the total scattering cross section $\sigma$, and the total LDOS, all are plotted versus $E$. Obviously, there is a good correspondence between the peaks of the total LDOS resulted from QBSs of different angular momenta (labelled TLDOS) with those of the total (labelled $\sigma$) and transverse (labelled $\eta$) scattering cross sections. Note that the (rather shallow) peaks of the transverse scattering cross section come both as maxima and minima. 

\section{Conclusion}
\label{sec: conclusion}
We have developed the $\T$-matrix formalism for studying electronic properties of the GQDs induced by a cylindrically symmetric confinement potential (circular GQDs). It was first shown that for circular GQDs with any radial confinement potential the equations for the bound states and QBSs spectra as well as the associated quantities such as the LDOS or scattering coefficients are all expressed explicitly in terms of the corresponding $\T$-matrix. In the case of simple confinement potentials (e.g., rectangular one), when the Dirac-like equation can be solved analytically, these equations give exactly the analytical results reported in various references. For any complicated potential, the $\T$-matrix can be determined numerically. As an example, we have in detail considered the case of trapezoidal radial confinement potentials, calculating the bound states and QBSs spectra, the LDOS, the differential scattering cross section, and the transverse scattering cross section for the potentials of different parameters. Apart from the role of a demonstration for the $\T$-matrix approach developed, obtained results in this example, in particular, suggest that controlling the valley polarization of Dirac electrons may turn out to be difficult in the presence of electrostatic disorders and/or thermal fluctuation. As an addition, we have shown how the developed $\T$-matrix formalism can be extended to study circular GQDs under a homogeneous perpendicular magnetic field (Appendix~\ref{sec: with B}). 
\acknowledgements 
We thank Cong Huy Pham and Duy Quang To for useful discussions.   
This work was supported by Vietnam National Foundation for Science and Technology Development (NAFOSTED) under Grant No. 103.02-2013.17.
\appendix
\section{$\W$-matrix for $E = \bar{U} \pm \nu \Delta$}
\label{sec: zero-energy}
As was mentioned above, the $\W (\bar{U},r)$-matrix as defined in (\ref{eq: W-matrix}) diverges as $E \rightarrow \bar{U} \pm \nu\Delta$. Note that the basic solutions are always defined up to constant factors that do not depend on $r$. To cure this divergence, we introduce the regularization factors to the basic solutions so that they remain finite in the limit of $E \rightarrow \bar{U} \pm \nu \Delta$. For example, for $j > 0$, the regularized $\W$-matrix can be defined as
\begin{equation}
\tilde{\W} (\bar{U} ,r) = 
\left(
	\begin{array}{cc}
	\tilde{J}_{j-\frac{1}{2}} (q r) & -i \epsilon^{(+)} \tilde{Y}_{j-\frac{1}{2}} (q r) \\
	i \epsilon^{(-)} \tilde{J}_{j+\frac{1}{2}} (q r) &  \tilde{Y}_{j+\frac{1}{2}} (q r)
	\end{array}
\right),
\label{eq: W-matrix 1}
\end{equation}
where $\tilde{J}_{j\pm\frac{1}{2}} (q r) = q^{-\abs{j\pm\frac{1}{2}}}J_{j\pm\frac{1}{2}} (q r)$, $\tilde{Y}_{j\pm\frac{1}{2}} (q r) = q^{\abs{j\pm\frac{1}{2}}}Y_{j\pm\frac{1}{2}} (q r)$ and $\epsilon^{(\pm)} = E - \bar{U} \pm \nu \Delta$. Now, letting $E \rightarrow \bar{U} \pm \nu \Delta$, we find
\begin{equation}
\tilde{\W} (\bar{U} ,r) \rightarrow
\left(
	\begin{array}{cc}
	r^{\abs{j-\frac{1}{2}}} & - \frac{i\epsilon^{(+)}}{2j-1} r^{-\abs{j-\frac{1}{2}}} \\
	\frac{i\epsilon^{(-)}}{2j+1} r^{\abs{j+\frac{1}{2}}} & r^{-\abs{j+\frac{1}{2}}}
	\end{array}
\right),
\label{eq: W-matrix 1 degenerate 1}
\end{equation}
for $j>\frac{1}{2}$, where we have also removed common constant factors in taking the limit. For $j=\frac{1}{2}$, the limit is more tricky, where one also needs to linearly recombine the two solutions to find 
\begin{equation}
\tilde{\W} (\bar{U} ,r) \rightarrow
\left(
\begin{array}{cc}
1 & i \epsilon^{(+)} \ln r \\
\frac{i \epsilon^{(-)}}{2} r & \frac{1}{r}
\end{array}
\right).
\label{eq: W-matrix 1 degenerate 2}
\end{equation}
It is easy to check that the wave functions of Eq.~(\ref{eq: W-matrix 1 degenerate 1})  and Eq.~(\ref{eq: W-matrix 1 degenerate 2}) are really the solutions to the Dirac equation~(\ref{eq: eigen chi U}) at $E = \bar{U} \pm \nu \Delta$. 

Similarly, for $j < 0$, we have
\begin{equation}
\tilde{\W} (\bar{U} , r) = 
\left(
	\begin{array}{cc}
	-i \epsilon^{(+)} \tilde{J}_{j-\frac{1}{2}} (q r) & \tilde{Y}_{j-\frac{1}{2}} (q r) \\
	 \tilde{J}_{j+\frac{1}{2}} (q r) &  i \epsilon^{(-)} \tilde{Y}_{j+\frac{1}{2}} (q r)
	\end{array}
\right).
\label{eq: W-matrix 2}
\end{equation}
Using the same procedure of taking the limit $E \rightarrow \bar{U} \pm \nu \Delta$ as above, we finds
\begin{eqnarray}
\tilde{\W} (\bar{U} ,r) \rightarrow 
\left(
	\begin{array}{cc}
	-\frac{i\epsilon^{(+)}}{2j-1} r^{\abs{j-\frac{1}{2}}} & r^{-\abs{j-\frac{1}{2}}}   \\
	r^{\abs{j+\frac{1}{2}}} & \frac{i \epsilon^{(-)}}{2j+1} r^{-\abs{j+\frac{1}{2}}}
	\end{array}
\right),
\label{eq: W-matrix 2 degenerate 1}
\end{eqnarray}
for $j < - \frac{1}{2}$, and
\begin{eqnarray}
\tilde{\W} (\bar{U} ,r) \rightarrow 
\left(
\begin{array}{cc}
\frac{i \epsilon^{(+)}}{2} r & \frac{1}{r}   \\
1 & i \epsilon^{(-)} \ln r
\end{array}
\right),
\label{eq: W-matrix 2 degenerate 2}
\end{eqnarray}
for $j=-\frac{1}{2}$.

Note that, by definition, $\T$-matrix is basis-dependent. So, in the limiting case studied, when the $\W$-matrix of Eq.~(\ref{eq: W-matrix}) is replaced by $\tilde{\W}$ defined above, all the boundary conditions and the Eqs.~(\ref{eq: bound states}), (\ref{eq: quasi-bound states}), (\ref{eq: dos}) and (\ref{eq: scattering coefficients}) should be slightly modified accordingly. 

Actually, the discussed degenerate solution is responsible for a special kind of bound states when the potential satisfies certain conditions. Since these special states only exist under very particular conditions, we do not examine them in any detail and interested readers are referred to Refs.~\cite{Hewageegana2008a,Downing2011a}. 
\section{$\T$-matrix for circular GQDs in a magnetic field}
\label{sec: with B}

In the presence of a uniform magnetic field $B$, the Hamiltonian of Eq.~(\ref{eq: H U}) becomes 
\begin{equation}
H_{\tau}= v_F \vec{\sigma} (\vec{p}+\frac{e}{c} \vec{A}) + \nu \Delta \sigma_z + U(r),
\label{eq: H B}
\end{equation}
where $\vec{A}$ is the vector potential~\cite{Recher2009a}. Note that we explicitly reintroduce in this Hamiltonian the Fermi velocity $v_F$ and the Planck constant $\hbar$ to distinguish the scale of quasi-relativistic effects (defined by $v_F$) and the scale of electrodynamics (defined by $c$). The magnetic field is assumed to be perpendicular, $\vec{B} = (0, 0, B)$, and we choose the symmetric gauge,  $\vec{A}=\frac{B}{2}(-y, x, 0)$. It is well-known that perpendicular magnetic field can induce localization of Dirac electrons even in the absence of the band gap~\cite{Apergel2010a}. In fact, for strong magnetic field, Dirac electrons are expected to exhibit the relativistic Landau levels~\cite{Neto2009a}. The effects of weak and medium magnetic field on the electron localization in electrostatic GQDs have been also discussed early~\cite{Chen2007a,Giavaras2009a}. The spectral equation for a rectangular GQD with a perpendicular magnetic field can be written down explicitly~\cite{Giavaras2009a,Masir2009a}. We will show that for a general electrostatic potential of the form~(\ref{eq: general potential}), the spectral equation can also be written in terms of the $\T$-matrix with some modification. 

Since the magnetic field preserves the cylindrical symmetry of the system, the Hamiltonian (\ref{eq: H B}) can be dealt with in terms of the $\T$-matrix in the same way as that described in Sec.~\ref{sec: general}. Indeed, using the ansatz (\ref{eq: eigen j}) for the eigenvalue problem of the Hamiltonian (\ref{eq: H B}), we obtain the equation for the radial spinor $\chi=(\chi_A,\chi_B)^t$ as
\begin{widetext}
\begin{equation}
	\left(
	\begin{array}{cc}
	U(r) - E + \nu \Delta  & - i\hbar v_F \left(\partial_r + \frac{j+\frac{1}{2}}{r} + \frac{r}{2l_B^2} \right) \\
	- i\hbar v_F \left(\partial_r - \frac{j-\frac{1}{2}}{r} - \frac{r}{2 l_B^2} \right) & U(r) -E - \nu \Delta
	\end{array}
	\right)
	\left(
	\begin{array}{c}
	\chi_{A} (r) \\
	\chi_{B} (r)
	\end{array}
	\right)
	= 0,
\label{eq: chi A B}
\end{equation}
\end{widetext}
where $l_B$ is the magnetic characteristic length, $l_B=\sqrt{{\hbar c}/{eB}}$.

Again, we consider the Eq.~(\ref{eq: chi A B}) in some region $r_a < r < r_b$ where the potential is constant, $U(r) = \bar{U}$. Following Ref.~\cite{Recher2009a}, the general solution to this equation can be written in terms of the Kummer functions $U$ (not to be confused with the potential) and $M$~\cite{Abramowitz1972a},
\begin{eqnarray}
\chi (r)=&& e^{-br^2/2} r^{n_\sigma} \left[ C^{(1)} \alpha_\sigma M (q_\sigma,1+n_\sigma,br^2) \right. \nonumber \\
&&\qquad \left. + C^{(2)}\beta_\sigma U (q_\sigma,1+n_\sigma,br^2)\right],
\end{eqnarray}
where $q_{\sigma}= \frac{1}{4}\left[\frac{a_{\sigma }}{b}+2(1+n_\sigma) \right]$,  $a_{\sigma } = 2 b \left(j+\frac{\sigma}{2}\right)- [(E - \bar{U})^2 - \Delta^2 ] /(\hbar v_F)^2$ and $n_{\sigma}=\left| j - \frac{\sigma}{2} \right|$ ($\sigma=A/B$ is identified with $\sigma=\pm 1$), $b=1/2l_B^2$. The coefficients  $\alpha_\sigma$ and $\beta_\sigma$ are defined only up to their relative ratios, which are $\frac{\alpha_{-}}{\alpha_{+}}=2bi\frac{\hbar v_F}{E-\bar{U} +\nu \Delta}\left(1-\frac{q_{+}}{1+n_{+}}\right)$, $\frac{\beta_{-}}{\beta_{+}}=2bi\frac{\hbar v_F}{E - \bar{U} + \nu \Delta}$ for $j>0$; and   $\frac{\alpha_{-}}{\alpha_{+}} = i\frac{E - \bar{U} -\nu \Delta}{\hbar v_F}\frac{1+n_{-}}{2bq_{-}}$, $\frac{\beta_{-}}{\beta_{+}}= - i\frac{E-\bar{U} - \nu \Delta}{\hbar v_F}\frac{1}{2bq_{-}}$ for $j<0$. This solution can be written in the form $\chi (r) = \W (\bar{U} , r) C$, with $C = (C^{(1)}, C^{(2)})^t$ and
\begin{widetext}
\begin{equation}
\W(\bar{U}, r) = e^{-\frac{br^2}{2}} \left(
\begin{array}{cc}
 \alpha_{+}  r^{n_{+}} M (q_{+},1+n_{+},br^2) &  \beta_{+}  r^{n_{+}} U (q_{+},1+n_{+},br^2) \\
 \alpha_{-}  r^{n_{-}} M (q_{-},1+n_{-},br^2) &  \beta_{-}  r^{n_{-}} U (q_{-},1+n_{-},br^2)
\end{array}
\right). 
\end{equation}
\end{widetext}
Further, viewing $C=(C^{(1)},C^{(2)})^t$ as the local basic coefficients we can consider an arbitrary radial potential of the form (\ref{eq: general potential}) and follow the $\T$-matrix formalism just developed in this paper. 

In particular, $\T$-matrix can be defined as the matrix that maps the basic coefficients $C_i$ in the limiting region of small $r$ to the basic coefficients $C_f$ in the limiting region of large $r$.  The bound states in a circular GQD under a perpendicular magnetic field can be identified as follows. Since the Kummer function $U$ is singular at the origin~\cite{Abramowitz1972a}, the basic coefficients near the origin should have a vanishing component associated with $U$, $C_i \propto (1,0)^{t}$. On the other hand, in the limiting region of large $r$, the Kummer function $M$ is singular~\cite{Abramowitz1972a}, and should not be present in the basic coefficients in this region, implying $C_f \propto (0,1)^{t}$. As a result, the spectral equation for bound states of a circular GQD under a uniform magnetic field reads
\begin{equation}
\T_{11}= 0.
\end{equation}
For a rectangular potential this equation reduces to the spectral equations reported in Refs.~\cite{Recher2009a,Rubio2015a}. Perpendicular magnetic fields may induce significant effects such as $(i)$ enhancing the localization of QBSs, $(ii)$ creating new bound states, and $(iii)$ lifting the valley degeneracy \cite{Recher2009a,Chen2007a}. For a negative angular momentum a perpendicular magnetic field can even induce the localisation-delocalisation-localisation transition \cite{Chen2007a}. Particularly, the truly bound states as those in conventional semiconductor quantum dots can in principle be created by a spatially non-uniform magnetic field \cite{deMartino2007a}. 

\bibliography{hcnguyen}

\end{document}